\documentclass[aps,twocolumn,groupedaddress,showpacs,showkeys,nofootinbib,amsfonts,eqsecnum,floatfix]{revtex4-1}

\usepackage[utf8]{inputenc}
\usepackage{amsmath}
\usepackage{amsfonts}
\usepackage{amssymb}
\usepackage[usenames,dvipsnames]{color}
\usepackage{pstricks}
\usepackage{soulutf8}
\usepackage{datetime}
\usepackage{hyphenat}
\usepackage{marginnote}
\usepackage{commath}
\usepackage{graphicx}

\newcommand{\Eqref}[1]{eq.~\eqref{#1}}
\newcommand{\Eq}[2]{
\begin{equation}
#2
\label{#1}
\end{equation}
}
\newcommand{\Eqs}[2]{
\label{#1}
\begin{align}
#2
\end{align}
}
\newcommand{\Fig}[3]{
\begin{figure}[]
	\centering
	\includegraphics{#1}
	\caption{#3}
	\label{#2}
\end{figure}
}
\newcommand{\Figref}[1]{fig.~\ref{#1}}

\newcommand{\KO}{\mathcal{O}}

\newcommand{\KF}{\mathcal{F}}
\newcommand{\KVol}{\mathrm{Vol}(3)}

\newcommand{\KGeV}{\text{ GeV}}

\newcommand{\Kl}{\ell}

\newcommand{\KS}{\mathcal{S}}

\begin{document}

%%%%%%%%%%%%%%%%%%%%%%%%%%%%%%%%%%%%
%%%%%%%%%%%%%%%%%%%%%%%%%%%%%%%%%%%%
%%%%         Title              %%%%
%%%%%%%%%%%%%%%%%%%%%%%%%%%%%%%%%%%%
%%%%%%%%%%%%%%%%%%%%%%%%%%%%%%%%%%%%
\title{Trouble Finding the Optimal AdS/QCD}
%%%%%%%%%%%%%%%%%%%%%%%%%%%%%%%%%%%%
%%%%%%%%%%%%%%%%%%%%%%%%%%%%%%%%%%%%
%%%%          Authors           %%%%
%%%%%%%%%%%%%%%%%%%%%%%%%%%%%%%%%%%%
%%%%%%%%%%%%%%%%%%%%%%%%%%%%%%%%%%%%
\author{K. Veschgini}
\email{k.veschgini@tphys.uni-heidelberg.de}
\author{E. Meg\'{\i}as}
\email{emegias@tphys.uni-heidelberg.de}
\author{H.J. Pirner}
\email{pir@tphys.uni-heidelberg.de}

\affiliation{Institut für Theoretische Physik der Universität Heidelberg, Heidelberg, Germany}

%%%%%%%%%%%%%%%%%%%%%%%%%%%%%%%%%%%%
%%%%%%%%%%%%%%%%%%%%%%%%%%%%%%%%%%%%
%%%%          Abstract          %%%%
%%%%%%%%%%%%%%%%%%%%%%%%%%%%%%%%%%%%
%%%%%%%%%%%%%%%%%%%%%%%%%%%%%%%%%%%%
\begin{abstract}
In the bottom-up approach to AdS/QCD based on a five-dimensional
gravity dilaton action the exponential of the dilaton field is 
usually identified as the strong or 't Hooft coupling. There
is currently no model known which fits the measurements of 
the running coupling and lattice results for pressure at
the same time. With a one parametric toy model we demonstrate 
the effect of fitting the pressure on the coupling and vice versa.
\end{abstract}

%%%%%%%%%%%%%%%%%%%%%%%%%%%%%%%%%%%%
%%%%%%%%%%%%%%%%%%%%%%%%%%%%%%%%%%%%
%%%%        Introduction        %%%%
%%%%%%%%%%%%%%%%%%%%%%%%%%%%%%%%%%%%
%%%%%%%%%%%%%%%%%%%%%%%%%%%%%%%%%%%%



\maketitle

\section{Introduction}
In the bottom-up approach to AdS/QCD we modify the
Maldacena duality \cite{Maldacena:1997re} between
$\mathcal{N}=4$ super-Yang-Mills theory and Type IIB string theory
of $AdS_5\times S_5$ to find a holographic dual for QCD.
We break the conformal invariance by adding a non-trivial dilaton
potential $V(\phi)$ to the bulk action 
\cite{Gursoy:2007cb,Gursoy:2007er,Gursoy:2008za}
\Eq{eq:sbulk}{
\KS_\text{bulk} = \frac{-1}{16\pi G_5}\int
	\sqrt{G}\del{R-{4\over 3} \del{\partial_\mu\phi}^2-V(\phi)}
	\mathrm{d}^{5} x \,.
}
The five-sphere $S_5$ is of no importance for the purpose of this
paper. The integration is performed over Euclidean space-time
with periodic time axis and the bulk coordinate $z$. The main
challenge is to find the correct potential. One approach would be to
make directly an ansatz for $V$. Instead we will use 
$b_0(z)$, given by the zero temperature
solution of the Einstein equations
\Eq{eq:thermalGasMetric}{
	\dif s^2 = b_0^2(z)\del{\dif \tau^2 +  \dif \vec{x}\cdot
	 \dif \vec{x}+\dif z^2} \,, 
}
to define an energy scale. The $\beta$-function is then given by
\Eq{eq:defbeta}{
\beta(\alpha) = b_0 \dod{\alpha}{b_0} \, .
}
The running coupling $\alpha$ on the gauge side of 
the duality corresponds to the exponential of the dilaton field 
$\exp(\phi)$.
The $\beta$-function then fixes the potential which is obtained from the
zero temperature Einstein equations
\cite{Gursoy:2007cb,Gursoy:2007er,Galow:2009kw}
\Eqs{}{
	\partial_z W_0 &=\frac{16}{9}b_0 W_0^2 
	+\frac{3}{4}b_0 V(\alpha_0) \,, \label{eq:ei0w} \\
	\partial_z b_0 &=-\frac{4}{9}b_0^2 W_0\,,\label{eq:ei0b}\\
	\partial_z \alpha_0 &=\alpha_0\sqrt{b_0 \partial_z W_0}\,.\label{eq:ei0a}
}
namely\footnote{The minus signs in the dilaton potential and 
in front of $V(\phi)$ in \Eqref{eq:sbulk} are a matter of convention.}
\Eq{eq:vfrombeta}{
V(\alpha)=-\frac{12}{\Kl^2}\exp
	\del{-\frac{8}{9}
		\int\limits_0^\alpha 
		\frac{\beta(\tilde{\alpha})}{\tilde{\alpha}^2}	\dif \tilde{\alpha}
		}
		\del{1-\del{\frac{\beta(\alpha)}{3\alpha}}^2} \,.
}
$W_0$ is defined by \Eqref{eq:ei0b} to reduce Einstein equations to
first oder. Thus, up to the constant factor $12/\Kl^2$ the 
dilaton potential is fixed by the $\beta$-function.

A holographic model is meant to capture infrared physics
as, according to the AdS/CFT correspondence the gravity 
description of the super-Yang-Mills theory
applies to the large 't Hooft coupling limit 
$\lambda_t=g_\text{YM}^2 N_c \to \infty$. 
The ultraviolet behavior physics computed from 
a gravity dual is known to show often a wrong behavior.
For example, consider a holographic model with
the $\beta$-function
\Eq{}{\beta_\text{pert}(\alpha) =-\beta_0 \alpha^2 -\beta_1 \alpha^3 \,, }
in the ultraviolet. The asymptotic behavior of the
spatial string tension computed from the gravity dual in the limit $T\to \infty$
is $\sigma_s \propto T^2 \alpha^{4/3}$ \cite{Alanen:2009ej,Megias:2010ku} instead of 
$\sigma_s \propto T^2 \alpha^2$ as it follows from dimensional reduction
arguments and lattice simulations \cite{Boyd:1996bx,Lucini:2002wg}. 
Similarly if we compare the asymptotic behavior of the pressure
\cite{Megias:2010tj,Megias:2010ku}
\Eq{eq:pasym}{ 
\begin{split}
p &= \frac{\pi^3 \Kl^3}{16 G_5} T ^4\del{
1-\frac{4}{3}\beta_0 \alpha_h
+\frac{2}{9}\del{4\beta_0^2-3\beta_1} \alpha_h^2
}  \\
 &= p_\text{SB} \del{
1-2.33 \alpha_h
+1.86 \alpha_h^2
} \,, 
\end{split}
}
with the perturbative result from QCD \cite{Kajantie:2002wa}
\Eq{eq:ppertqcd}{ 
\begin{split}
p&=\frac{8\pi^2}{45}T^4
\del{
1-\frac{15}{4} \frac{\alpha}{\pi}
+30 \del{\frac{\alpha}{\pi}}^{3/2}
} \\
&=p_\text{SB}\del{
1-1.19 \alpha
+5.4\alpha^{3/2}
}\,,
\end{split}
}
we see that not only the coefficients are different but also  the power of $\alpha$
in the next to leading order. 
Note, pertubation theory gives the coupling at the black horizon $\alpha_h=\alpha(z_h)$
equal to $\alpha(\pi T)$ in QCD in lowest order.
In many cases, like the spatial tension,
the ultraviolet limit does not prevent the model from capturing the 
infrared physics, but in the case of the pressure the situation is different. The
gravity model suggests a smaller pressure than perturbative QCD at
very high temperatures, as we can see from a comparison of the coefficients at 
$\KO(\alpha)$ in \Eqref{eq:pasym} and \Eqref{eq:ppertqcd}.
If the pressure is already much too small at $10^3 T_c$, it has a tendency 
to be also much too small at lower temperatures. 

We will consider a simple $\beta$-function and demonstrate that
as we switch over at smaller values of $\hat{\alpha}$ 
from $\beta_\text{pert}$ to an asymptotic linear
behavior the pressure gets larger. Fitting lattice results requires
$\hat{\alpha} \lesssim 0.04$.

%%%%%%%%%%%%%%%%%%%%%%%%%%%%%%%%%%%%
%%%%%%%%%%%%%%%%%%%%%%%%%%%%%%%%%%%%
%%%%      The Model             %%%%
%%%%%%%%%%%%%%%%%%%%%%%%%%%%%%%%%%%%
%%%%%%%%%%%%%%%%%%%%%%%%%%%%%%%%%%%%

\section{The model}
We assume the following toy $\beta$-function to demonstrate 
our case:
\Eq{Eq:betaSimple}{
\beta(\alpha)= \begin{cases}
	\beta_\text{pert}(\alpha)-\hat{\beta}_2 \alpha^4 & \text{ if } \alpha 
	\le \hat{\alpha} \\
	\beta_\text{pert}(\hat{\alpha})-\hat{\beta}_2 \hat{\alpha}^4 
	-3 (\alpha- \hat{\alpha}) & 
	\text{ if } \alpha > \hat{\alpha}
	\end{cases}\,,
	}
where
\Eq{}{
	\hat{\beta}_2 = 
	\frac{3-2\beta_0 \hat{\alpha}-3 \beta_1 \hat{\alpha}^2}
	{4\hat{\alpha}^3} \,,
}
is chosen such that $\partial_\alpha \beta(\alpha)$ is continuous.
There is only one parameter $\hat{\alpha}$ which controls the
transition point. Note that $\hat{\beta}_2$ will become very large
for small $\hat{\alpha}$ such that $\beta(\alpha)$ deviates
from $\beta_\text{pert}(\alpha)$ already at $\alpha$ much smaller
than $\hat{\alpha}$. The slope of the linear term is chosen to be
$-3$. With this choice, according to \Eqref{eq:vfrombeta} we
obtain a monotonic potential with the following asymptotic
behaviors:
\begin{alignat}{2}
V(\alpha)&  \to -12/\Kl^2 \;&;&\; \alpha\to0 \,, \\
V(\alpha)&  \sim -\alpha^{5/3} &;&\; \alpha\to\infty \,.
\end{alignat}

\section{The pressure}

\Fig{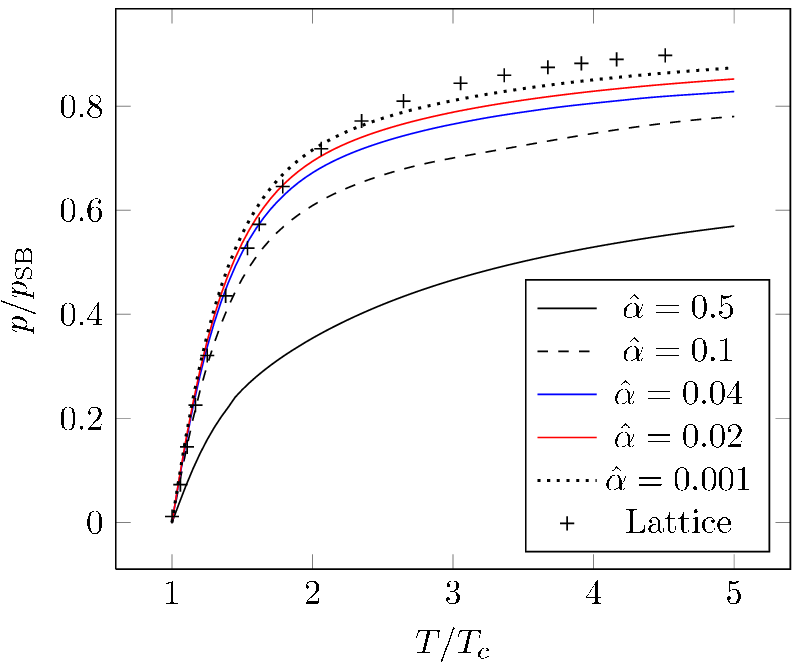}{fig:p-lowT}{Pressure normalized to the Stefan-Boltzmann 
pressure as a function of $T/T_c$
compared with lattice data for $N_c=3$ taken from ref.~\cite{Boyd:1996bx}.}

The deconfined phase of the gluon plasma is described in the
holographic picture by a black hole geometry with the metric
\Eq{}{
	\dif s^2 = b^2(z)\del{ f(z) \dif \tau^2 +  \dif \vec{x}\cdot 
	\dif \vec{x} +{ \dif z^2 \over f(z) }} \,.
}
The horizon lies in the bulk at $z_h$ where $f(z_h)=0$.
We normalize $f(0)$ to 1. In order to avoid a conical singularity
the periodicity of the $\tau$ axes $\beta=T^{-1}$ must be fixed to
\cite{Carlip:2008wv,PhysRevD.15.2752}
\Eq{}{
\beta = -\eval{\frac{4 \pi}{\partial_z f(z)}}_{z=z_h} \,.
} 
The zero temperature solution
serves as background. Any finite temperature solution
shares the same ultraviolet behavior as the zero temperature
solution up to $\KO(z^4)$. The physical action is given by
the difference between the action of the black hole solutions
and the zero temperature solution
\Eq{}{
\KS_\text{phys.} = \KS_T - \KS_0 \,,
}
where $\KS_T$ and $\KS_0$ are actions given by the
bulk term \Eqref{eq:sbulk} plus the Gibbons-Hawking-York boundary term 
\cite{PhysRevD.15.2752,PhysRevLett.28.1082}
\Eq{}{
\KS_\text{bound.}= {1 \over 8 \pi G_5} \int \dif x^4\sqrt{g} K \,.
}
The induced metric on the boundary is $g$ and $K$ is 
the trace of the second fundamental form of the boundary. 
For temperatures $T$ larger than some 
$T_\text{min}$ we find three 
solutions, the zero temperature
solution (temperature independent) and two black hole solutions.
For $T<T_\text{min}$ the black hole solutions are not present.
The solution with the smaller physical action defines the stable
gluonic matter. The phase of confined gluons is given
from $T=0$ up to $T=T_c$ by the zero temperature solution.
Black hole solutions exist for temperatures higher than some 
$T_\text{min}$, the big black holes have a smaller action than
the small black holes and for $T>T_c$ a negative physical action
(relative to the $T=0$ solution). The black hole geometry
corresponds to the deconfined phase.
Details of the computation can be found in ref.~\cite{Megias:2010ku}.
The free energy can be computed directly from the physical action
as $\KF = T \KS_\text{phys.} $ or by integrating the
entropy
\Eq{}
{
\KF = \int S \dif T \,.
}
The entropy of a classical black hole is a well defined quantity given
by the Bekenstein-Hawking formula  \cite{Bekenstein:1973ur,Hawking:1974sw}
\Eq{}
{
S = \frac{A}{4 G_5} = \frac{\KVol b^3(z_h)}{4 G_5}\,,
}
where $A$ is the area of the black hole horizon.
Both methods produce the same result but 
the latter approach is numerically favorable \cite{Megias:2010ku}.
Thermodynamic quantities can then be computed from the free energy.
The five-dimensional gravitational constant $G_5$ is fixed by
normalizing the pressure given in \Eqref{eq:pasym} to the Stefan-Boltzmann 
pressure in the limit $T\to\infty$:
\Eq{}{
G_5 = \frac{45\pi \Kl^3}{16 (N_c^2-1)} \,.
}
This value differs from the conformal case because
we have fewer degrees of freedom in QCD
than in $\mathcal{N}=4$ super Yang-Mills theory 
\cite{Schafer:2009dj}.
We have computed the pressure using Bekenstein-Hawking entropy for
different choices of $\hat{\alpha}$. The results normalized to the 
ideal gas limit are shown in \Figref{fig:p-lowT}. 
Lattice data from $T_c$ up to $2 T_c$ are fitted very well 
for $\hat{\alpha}=0.04$. The choice $\hat{\alpha}=0.02$
gives the best fit from $T_c$ up to $3 T_c$. Some other curves are
also shown for comparison. Data above $3 T_c$ are not fitted
precisely by any choice of $\hat{\alpha}$.

%%%%%%%%%%%%%%%%%%%%%%%%%%%%%%%%%%%%
%%%%%%%%%%%%%%%%%%%%%%%%%%%%%%%%%%%%
%%%%      The running coupling  %%%%
%%%%%%%%%%%%%%%%%%%%%%%%%%%%%%%%%%%%
%%%%%%%%%%%%%%%%%%%%%%%%%%%%%%%%%%%%

\section{The running coupling and screening mass}
\label{sec:coupling}

It is interesting to see how the running coupling $\alpha$
is affected by the choice of $\hat{\alpha}$. The 
weak coupling expansion \Eqref{eq:pasym} already suggests that
the coupling must be very small when we want to fit the pressure.
First we discuss the zero temperature running coupling
as a function of energy $E$, which we denote by $\alpha_E$.
Since we define the $\beta$-function with $b_0$ 
as given in \Eqref{eq:defbeta}, the
energy is proportional to $b_0(z)$. The proportionality constant 
$\Lambda_E$ is not fixed by the model. Different values of $\Lambda_E$ 
correspond to different initial conditions
for the renormalization group equation \eqref{eq:defbeta}. Fig.~\ref{fig:aE} shows $\alpha_E$ as 
a function of energy $E$ in logarithmic scale. Different 
values of $\Lambda_E$ shift the curves to the left or right without
affecting their shape since
$\log(E=\Lambda_E b_0)=\log(\Lambda_E)+\log(b_0)$. 
This is demonstrated for the case 
$\hat{\alpha}=0.02$, where we fix $\Lambda_E$ either by 
\Eq{}{
\alpha_E(1.78\text{GeV})=0.33
}
or by
\Eq{}{
\alpha_E(M_Z=91.2\text{GeV})=0.118
}
according to 
ref.~\cite{Bethke:2009jm}.
A small value of $\hat{\alpha}$
results in a very small coupling at high energies. As a side effect the 
curve becomes also very flat, there is almost no running at high energies.
On the other hand the coupling 
rises extremely fast in the infrared. This differs from the
case $\hat{\alpha}=0.5$ which is a much better fit to the 
$\overline{\text{MS}}$ value in the plotted range.

\begin{figure}
\centering
\includegraphics[width=7.8cm,height=6.8cm]{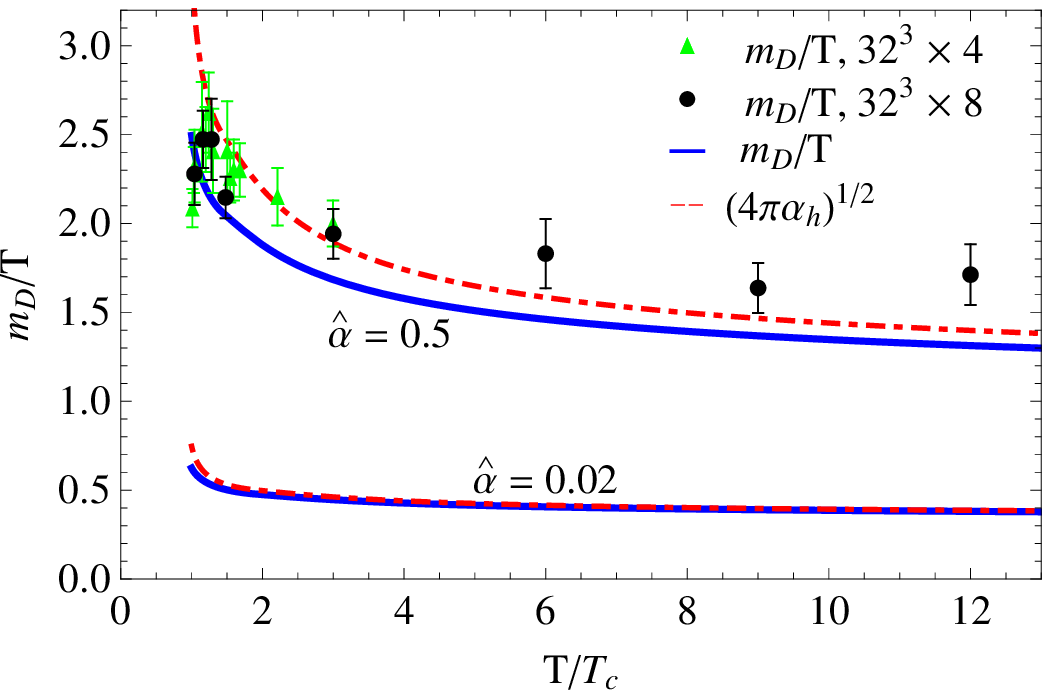}
\caption{Debye mass over temperature as a function of $T/T_c$.  The
  full (blue) curves correspond to computations using
  $\hat{\alpha}=0.02$ and $\hat{\alpha}=0.5$. Lattice data are taken
  for gluodynamics SU(3) with $N_\sigma^3 \times N_\tau = 32^3 \times
  4$ and $32^3\times 8$ from
  refs.~\cite{Kaczmarek:2004gv,Kaczmarek:2005ui}.  The dot dashed
  (red) curves represent the finite temperature running coupling
  $(4\pi \alpha(z_h))^{1/2}$ for both cases.}
\label{fig:mD}
\end{figure}

The chosen $\beta$-function for $\hat{\alpha}=0.5$ can reproduce the
running coupling in vacuum quite well. The coupling at finite
temperatures defined as $\alpha_h(T)=\alpha(z_h)$ follows the vacuum
coupling constant in the ultraviolet
\cite{Gursoy:2008za,Megias:2010ku}.  Thus~$\alpha_h$ is also very
small at high temperatures.  In order to test the model at finite
temperatures $T$, it is important to monitor the behavior of an
observable which is closely related to the coupling at finite
temperatures. To this end we have computed the Debye mass in the gluon
plasma.  We use the method presented in ref.~\cite{Gorsky:2010xf}. The
results for $\hat{\alpha} = 0.02$ and $\hat{\alpha} = 0.5 $ are
plotted in~\Figref{fig:mD} together with lattice
data~\cite{Kaczmarek:2004gv,Kaczmarek:2005ui}.  The model with
$\hat{\alpha} = 0.02$ predicts a value for the Debye mass which is a
factor 4 smaller than lattice data. This is a consequence of the fact
that the running coupling $\alpha_h$ is very small at high temperature
for this choice of $\hat{\alpha}$.  For the large $\hat{\alpha} = 0.5$
the Debye mass is quite close to the lattice data as one can see in
\Figref{fig:mD}.

\Fig{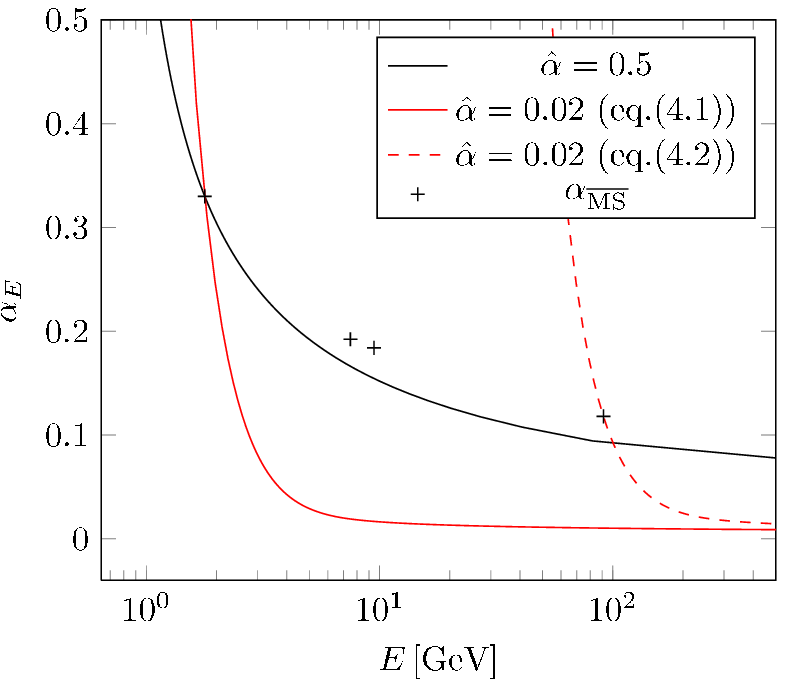}{fig:aE}{The running coupling $\alpha$ as a function
energy $E=\Lambda_E b_0$. Data points are taken from ref. ~\cite{Bethke:2009jm}.
The curve corresponding to $\hat{\alpha}=0.02$ is shown for two 
different choices of $\Lambda_E$.}

%%%%%%%%%%%%%%%%%%%%%%%%%%%%%%%%%%%%
%%%%%%%%%%%%%%%%%%%%%%%%%%%%%%%%%%%%
%%%%    Conclution              %%%%
%%%%%%%%%%%%%%%%%%%%%%%%%%%%%%%%%%%%
%%%%%%%%%%%%%%%%%%%%%%%%%%%%%%%%%%%%
\section{Summary and conclusion}

The $\beta$-function, on the one hand defines the running of the
coupling and on the other hand determines the dilaton potential and
hence the full thermodynamics including the pressure and the screening mass. 
Following the
idea of AdS/CFT we identify the exponential of the dilaton field as
the running coupling of QCD. But, when we fit the pressure, the
running coupling $\alpha$ and the screening mass in the plasma~$m_D$
deviate strongly from phenomenology. On the other side, if we choose
to fit $\alpha$ or $m_D$, we obtain a pressure that is about 40\% too small.
This outcome is not just a property of the model introduced here cf.
\cite{Gursoy:2009jd,Megias:2010tj,Alanen:2009xs}.  In the literature
refs.~\cite{Gursoy:2009jd,Megias:2010tj} both approaches have been
covered. In ref.~\cite{Gursoy:2009jd} a very good description of the
thermodynamics has been achieved, however at the expense of not being
able to relate the dilaton potential with the running $\alpha(E)$ in
the $\overline{\text{MS}}$ scheme. Note, one could argue, that in the
scheme used here with a large coefficient $\hat{\beta}_2 \alpha^4$ in
the $\beta$-function, the $\overline{\text{MS}}$-value of
$\alpha_{\overline{\text{MS}}}(E)=0.33$ at $E=1.78\KGeV$ is not
realistic. But as shown in section \ref{sec:coupling} any other choice
leads to the same result.  
Also the second thermodynamic observable, the Debye mass, points to a real deficit for
this choice of small $\hat \alpha$. 
For our toy model as well as for the model
of ref.~\cite{Gursoy:2009jd} there is currently no known mapping
between the assumed $\beta$-function and known $\overline{\text{MS}}$
values of $\alpha$ \cite{Gursoy:2010fj}. This would be very much
needed if one wants to apply this model to hadronization.  In
ref.~\cite{Megias:2010tj} an extrapolation of the $\beta$-function was
used which agrees well the perturbative running of
$\alpha_{\overline{\text{MS}}}$ \cite{Galow:2009kw}. The calculations
of the thermodynamics however have the default shown by the toy-model
for $\hat{\alpha}\approx 0.5$. It produces a pressure which is at all
temperatures too small, whereas the calculation of the Debye mass comes out in
agreement with the lattice data. At the moment we do not know any solution of
this dilemma.

\begin{acknowledgments}
E.M. would like to thank the Humboldt Foundation for their
stipend. This work was also supported in part by the ExtreMe Matter
Institute EMMI in the framework of the Helmholtz Alliance Program of
the Helmholtz Association.
\end{acknowledgments}

\bibliography{main}

%Merlin.mbs v4.21 2009-07-09.
\begin{thebibliography}{10}%
\makeatletter
\providecommand \@ifxundefined [1]{%
 \ifx #1\undefined \expandafter \@firstoftwo
 \else \expandafter \@secondoftwo
\fi
}%
\providecommand \@ifnum [1]{%
 \ifnum #1\expandafter \@firstoftwo
 \else \expandafter \@secondoftwo
\fi
}%
\providecommand \enquote [1]{``#1''}%
\providecommand \bibnamefont  [1]{#1}%
\providecommand \bibfnamefont [1]{#1}%
\providecommand \citenamefont [1]{#1}%
\providecommand\href[0]{\@sanitize\@href}%
\providecommand\@href[1]{\endgroup\@@startlink{#1}\endgroup\@@href}%
\providecommand\@@href[1]{#1\@@endlink}%
\providecommand \@sanitize [0]{\begingroup\catcode`\&12\catcode`\#12\relax}%
\@ifxundefined \pdfoutput {\@firstoftwo}{%
 \@ifnum{\z@=\pdfoutput}{\@firstoftwo}{\@secondoftwo}%
}{%
 \providecommand\@@startlink[1]{\leavevmode\special{html:<a href="#1">}}%
 \providecommand\@@endlink[0]{\special{html:</a>}}%
}{%
 \providecommand\@@startlink[1]{%
  \leavevmode
  \pdfstartlink
   attr{/Border[0 0 1 ]/H/I/C[0 1 1]}%
   user{/Subtype/Link/A<</Type/Action/S/URI/URI(#1)>>}%
  \relax
 }%
 \providecommand\@@endlink[0]{\pdfendlink}%
}%
\providecommand \url  [0]{\begingroup\@sanitize \@url }%
\providecommand \@url [1]{\endgroup\@href {#1}{\urlprefix}}%
\providecommand \urlprefix [0]{URL }%
\providecommand \Eprint[0]{\href }%
\@ifxundefined \urlstyle {%
  \providecommand \doi [1]{doi:\discretionary{}{}{}#1}%
}{%
  \providecommand \doi [0]{doi:\discretionary{}{}{}\begingroup
  \urlstyle{rm}\Url }%
}%
\providecommand \doibase [0]{http://dx.doi.org/}%
\providecommand \Doi[1]{\href{\doibase#1}}%
\providecommand \bibAnnote [3]{%
  \BibitemShut{#1}%
  \begin{quotation}\noindent
    \textsc{Key:}\ #2\\\textsc{Annotation:}\ #3%
  \end{quotation}%
}%
\providecommand \bibAnnoteFile [2]{%
  \IfFileExists{#2}{\bibAnnote {#1} {#2} {\input{#2}}}{}%
}%
\providecommand \typeout [0]{\immediate \write \m@ne }%
\providecommand \selectlanguage [0]{\@gobble}%
\providecommand \bibinfo [0]{\@secondoftwo}%
\providecommand \bibfield [0]{\@secondoftwo}%
\providecommand \translation [1]{[#1]}%
\providecommand \BibitemOpen[0]{}%
\providecommand \bibitemStop [0]{}%
\providecommand \bibitemNoStop [0]{.\EOS\space}%
\providecommand \EOS [0]{\spacefactor3000\relax}%
\providecommand \BibitemShut [1]{\csname bibitem#1\endcsname}%
%</preamble>
\bibitem{Maldacena:1997re}%
  \BibitemOpen
  \bibfield{author}{%
  \bibinfo {author} {\bibfnamefont{J.~M.}\ \bibnamefont{Maldacena}},\ }%
  \bibfield{journal}{%
  \Doi{10.1023/A:1026654312961}{\bibinfo {journal} {Adv. Theor. Math. Phys.}}\
  }%
  \textbf{\bibinfo {volume} {2}},\ \bibinfo {pages} {231} (\bibinfo {year}
  {1998}),\ \Eprint{http://arxiv.org/abs/hep-th/9711200}{arXiv:hep-th/9711200}%
  \bibAnnoteFile{NoStop}{Maldacena:1997re}%
%%CITATION = HEP-TH/9711200;%%
\bibitem{Gursoy:2007cb}%
  \BibitemOpen
  \bibfield{author}{%
  \bibinfo {author} {\bibfnamefont{U.}~\bibnamefont{Gursoy}}\ and\ \bibinfo
  {author} {\bibfnamefont{E.}~\bibnamefont{Kiritsis}},\ }%
  \bibfield{journal}{%
  \Doi{10.1088/1126-6708/2008/02/032}{\bibinfo {journal} {JHEP}}\ }%
  \textbf{\bibinfo {volume} {02}},\ \bibinfo {pages} {032} (\bibinfo {year}
  {2008}),\ \Eprint{http://arxiv.org/abs/0707.1324}{arXiv:0707.1324 [hep-th]}%
  \bibAnnoteFile{NoStop}{Gursoy:2007cb}%
%%CITATION = 0707.1324;%%
\bibitem{Gursoy:2007er}%
  \BibitemOpen
  \bibfield{author}{%
  \bibinfo {author} {\bibfnamefont{U.}~\bibnamefont{Gursoy}}, \bibinfo {author}
  {\bibfnamefont{E.}~\bibnamefont{Kiritsis}},\ and\ \bibinfo {author}
  {\bibfnamefont{F.}~\bibnamefont{Nitti}},\ }%
  \bibfield{journal}{%
  \Doi{10.1088/1126-6708/2008/02/019}{\bibinfo {journal} {JHEP}}\ }%
  \textbf{\bibinfo {volume} {02}},\ \bibinfo {pages} {019} (\bibinfo {year}
  {2008}),\ \Eprint{http://arxiv.org/abs/0707.1349}{arXiv:0707.1349 [hep-th]}%
  \bibAnnoteFile{NoStop}{Gursoy:2007er}%
%%CITATION = 0707.1349;%%
\bibitem{Gursoy:2008za}%
  \BibitemOpen
  \bibfield{author}{%
  \bibinfo {author} {\bibfnamefont{U.}~\bibnamefont{Gursoy}}, \bibinfo {author}
  {\bibfnamefont{E.}~\bibnamefont{Kiritsis}}, \bibinfo {author}
  {\bibfnamefont{L.}~\bibnamefont{Mazzanti}},\ and\ \bibinfo {author}
  {\bibfnamefont{F.}~\bibnamefont{Nitti}},\ }%
  \bibfield{journal}{%
  \Doi{10.1088/1126-6708/2009/05/033}{\bibinfo {journal} {JHEP}}\ }%
  \textbf{\bibinfo {volume} {05}},\ \bibinfo {pages} {033} (\bibinfo {year}
  {2009}),\ \Eprint{http://arxiv.org/abs/0812.0792}{arXiv:0812.0792 [hep-th]}%
  \bibAnnoteFile{NoStop}{Gursoy:2008za}%
%%CITATION = 0812.0792;%%
\bibitem{Galow:2009kw}%
  \BibitemOpen
  \bibfield{author}{%
  \bibinfo {author} {\bibfnamefont{B.}~\bibnamefont{Galow}}, \bibinfo {author}
  {\bibfnamefont{E.}~\bibnamefont{Megias}}, \bibinfo {author}
  {\bibfnamefont{J.}~\bibnamefont{Nian}},\ and\ \bibinfo {author}
  {\bibfnamefont{H.~J.}\ \bibnamefont{Pirner}},\ }%
  \bibfield{journal}{%
  \Doi{10.1016/j.nuclphysb.2010.03.022}{\bibinfo {journal} {Nucl. Phys.}}\ }%
  \textbf{\bibinfo {volume} {B834}},\ \bibinfo {pages} {330} (\bibinfo {year}
  {2010}),\ \Eprint{http://arxiv.org/abs/0911.0627}{arXiv:0911.0627 [hep-ph]}%
  \bibAnnoteFile{NoStop}{Galow:2009kw}%
%%CITATION = 0911.0627;%%
\bibitem{Alanen:2009ej}%
  \BibitemOpen
  \bibfield{author}{%
  \bibinfo {author} {\bibfnamefont{J.}~\bibnamefont{Alanen}}, \bibinfo {author}
  {\bibfnamefont{K.}~\bibnamefont{Kajantie}},\ and\ \bibinfo {author}
  {\bibfnamefont{V.}~\bibnamefont{Suur-Uski}},\ }%
  \bibfield{journal}{%
  \Doi{10.1103/PhysRevD.80.075017}{\bibinfo {journal} {Phys. Rev.}}\ }%
  \textbf{\bibinfo {volume} {D80}},\ \bibinfo {pages} {075017} (\bibinfo {year}
  {2009}),\ \Eprint{http://arxiv.org/abs/0905.2032}{arXiv:0905.2032 [hep-ph]}%
  \bibAnnoteFile{NoStop}{Alanen:2009ej}%
%%CITATION = 0905.2032;%%
\bibitem{Megias:2010ku}%
  \BibitemOpen
  \bibfield{author}{%
  \bibinfo {author} {\bibfnamefont{E.}~\bibnamefont{Megias}}, \bibinfo {author}
  {\bibfnamefont{H.~J.}\ \bibnamefont{Pirner}},\ and\ \bibinfo {author}
  {\bibfnamefont{K.}~\bibnamefont{Veschgini}}}%
   (\bibinfo {year} {2010}),\
  \Eprint{http://arxiv.org/abs/1009.2953}{arXiv:1009.2953 [hep-ph]}%
  \bibAnnoteFile{NoStop}{Megias:2010ku}%
%%CITATION = 1009.2953;%%
\bibitem{Boyd:1996bx}%
  \BibitemOpen
  \bibfield{author}{%
  \bibinfo {author} {\bibfnamefont{G.}~\bibnamefont{Boyd}} \emph{et~al.},\ }%
  \bibfield{journal}{%
  \Doi{10.1016/0550-3213(96)00170-8}{\bibinfo {journal} {Nucl. Phys.}}\ }%
  \textbf{\bibinfo {volume} {B469}},\ \bibinfo {pages} {419} (\bibinfo {year}
  {1996}),\
  \Eprint{http://arxiv.org/abs/hep-lat/9602007}{arXiv:hep-lat/9602007}%
  \bibAnnoteFile{NoStop}{Boyd:1996bx}%
%%CITATION = HEP-LAT/9602007;%%
\bibitem{Lucini:2002wg}%
  \BibitemOpen
  \bibfield{author}{%
  \bibinfo {author} {\bibfnamefont{B.}~\bibnamefont{Lucini}}\ and\ \bibinfo
  {author} {\bibfnamefont{M.}~\bibnamefont{Teper}},\ }%
  \bibfield{journal}{%
  \Doi{10.1103/PhysRevD.66.097502}{\bibinfo {journal} {Phys. Rev.}}\ }%
  \textbf{\bibinfo {volume} {D66}},\ \bibinfo {pages} {097502} (\bibinfo {year}
  {2002}),\
  \Eprint{http://arxiv.org/abs/hep-lat/0206027}{arXiv:hep-lat/0206027}%
  \bibAnnoteFile{NoStop}{Lucini:2002wg}%
%%CITATION = HEP-LAT/0206027;%%
\bibitem{Megias:2010tj}%
  \BibitemOpen
  \bibfield{author}{%
  \bibinfo {author} {\bibfnamefont{E.}~\bibnamefont{Megias}}, \bibinfo {author}
  {\bibfnamefont{H.~J.}\ \bibnamefont{Pirner}},\ and\ \bibinfo {author}
  {\bibfnamefont{K.}~\bibnamefont{Veschgini}},\ }%
  \bibfield{journal}{%
  \Doi{10.1016/j.nuclphysbps.2010.10.082}{\bibinfo {journal} {Nucl. Phys. Proc.
  Suppl.}}\ }%
  \textbf{\bibinfo {volume} {207-208}},\ \bibinfo {pages} {333} (\bibinfo
  {year} {2010}),\ \Eprint{http://arxiv.org/abs/1008.4505}{arXiv:1008.4505
  [hep-th]}%
  \bibAnnoteFile{NoStop}{Megias:2010tj}%
%%CITATION = 1008.4505;%%
\bibitem{Kajantie:2002wa}%
  \BibitemOpen
  \bibfield{author}{%
  \bibinfo {author} {\bibfnamefont{K.}~\bibnamefont{Kajantie}}, \bibinfo
  {author} {\bibfnamefont{M.}~\bibnamefont{Laine}}, \bibinfo {author}
  {\bibfnamefont{K.}~\bibnamefont{Rummukainen}},\ and\ \bibinfo {author}
  {\bibfnamefont{Y.}~\bibnamefont{Schroder}},\ }%
  \bibfield{journal}{%
  \Doi{10.1103/PhysRevD.67.105008}{\bibinfo {journal} {Phys. Rev.}}\ }%
  \textbf{\bibinfo {volume} {D67}},\ \bibinfo {pages} {105008} (\bibinfo {year}
  {2003}),\ \Eprint{http://arxiv.org/abs/hep-ph/0211321}{arXiv:hep-ph/0211321}%
  \bibAnnoteFile{NoStop}{Kajantie:2002wa}%
%%CITATION = HEP-PH/0211321;%%
\bibitem{Carlip:2008wv}%
  \BibitemOpen
  \bibfield{author}{%
  \bibinfo {author} {\bibfnamefont{S.}~\bibnamefont{Carlip}},\ }%
  \bibfield{journal}{%
  \Doi{10.1007/978-3-540-88460-6_3}{\bibinfo {journal} {Lect. Notes Phys.}}\ }%
  \textbf{\bibinfo {volume} {769}},\ \bibinfo {pages} {89} (\bibinfo {year}
  {2009}),\ \Eprint{http://arxiv.org/abs/0807.4520}{arXiv:0807.4520 [gr-qc]}%
  \bibAnnoteFile{NoStop}{Carlip:2008wv}%
%%CITATION = 0807.4520;%%
\bibitem{PhysRevD.15.2752}%
  \BibitemOpen
  \bibfield{author}{%
  \bibinfo {author} {\bibfnamefont{G.~W.}\ \bibnamefont{Gibbons}}\ and\
  \bibinfo {author} {\bibfnamefont{S.~W.}\ \bibnamefont{Hawking}},\ }%
  \bibfield{journal}{%
  \Doi{10.1103/PhysRevD.15.2752}{\bibinfo {journal} {Phys. Rev. D}}\ }%
  \textbf{\bibinfo {volume} {15}},\ \bibinfo {pages} {2752} (\bibinfo {month}
  {May}\ \bibinfo {year} {1977})%
  \bibAnnoteFile{NoStop}{PhysRevD.15.2752}%
\bibitem{PhysRevLett.28.1082}%
  \BibitemOpen
  \bibfield{author}{%
  \bibinfo {author} {\bibfnamefont{J.~W.}\ \bibnamefont{York}},\ }%
  \bibfield{journal}{%
  \Doi{10.1103/PhysRevLett.28.1082}{\bibinfo {journal} {Phys. Rev. Lett.}}\ }%
  \textbf{\bibinfo {volume} {28}},\ \bibinfo {pages} {1082} (\bibinfo {month}
  {Apr}\ \bibinfo {year} {1972})%
  \bibAnnoteFile{NoStop}{PhysRevLett.28.1082}%
\bibitem{Bekenstein:1973ur}%
  \BibitemOpen
  \bibfield{author}{%
  \bibinfo {author} {\bibfnamefont{J.~D.}\ \bibnamefont{Bekenstein}},\ }%
  \bibfield{journal}{%
  \Doi{10.1103/PhysRevD.7.2333}{\bibinfo {journal} {Phys. Rev.}}\ }%
  \textbf{\bibinfo {volume} {D7}},\ \bibinfo {pages} {2333} (\bibinfo {year}
  {1973})%
  \bibAnnoteFile{NoStop}{Bekenstein:1973ur}%
%%CITATION = PHRVA,D7,2333;%%
\bibitem{Hawking:1974sw}%
  \BibitemOpen
  \bibfield{author}{%
  \bibinfo {author} {\bibfnamefont{S.~W.}\ \bibnamefont{Hawking}},\ }%
  \bibfield{journal}{%
  \Doi{10.1007/BF02345020}{\bibinfo {journal} {Commun. Math. Phys.}}\ }%
  \textbf{\bibinfo {volume} {43}},\ \bibinfo {pages} {199} (\bibinfo {year}
  {1975})%
  \bibAnnoteFile{NoStop}{Hawking:1974sw}%
%%CITATION = CMPHA,43,199;%%
\bibitem{Schafer:2009dj}%
  \BibitemOpen
  \bibfield{author}{%
  \bibinfo {author} {\bibfnamefont{T.}~\bibnamefont{Schafer}}\ and\ \bibinfo
  {author} {\bibfnamefont{D.}~\bibnamefont{Teaney}},\ }%
  \bibfield{journal}{%
  \Doi{10.1088/0034-4885/72/12/126001}{\bibinfo {journal} {Rept. Prog. Phys.}}\
  }%
  \textbf{\bibinfo {volume} {72}},\ \bibinfo {pages} {126001} (\bibinfo {year}
  {2009}),\ \Eprint{http://arxiv.org/abs/0904.3107}{arXiv:0904.3107 [hep-ph]}%
  \bibAnnoteFile{NoStop}{Schafer:2009dj}%
%%CITATION = 0904.3107;%%
\bibitem{Bethke:2009jm}%
  \BibitemOpen
  \bibfield{author}{%
  \bibinfo {author} {\bibfnamefont{S.}~\bibnamefont{Bethke}},\ }%
  \bibfield{journal}{%
  \Doi{10.1140/epjc/s10052-009-1173-1}{\bibinfo {journal} {Eur. Phys. J.}}\ }%
  \textbf{\bibinfo {volume} {C64}},\ \bibinfo {pages} {689} (\bibinfo {year}
  {2009}),\ \Eprint{http://arxiv.org/abs/0908.1135}{arXiv:0908.1135 [hep-ph]}%
  \bibAnnoteFile{NoStop}{Bethke:2009jm}%
%%CITATION = 0908.1135;%%
\bibitem{Gorsky:2010xf}%
  \BibitemOpen
  \bibfield{author}{%
  \bibinfo {author} {\bibfnamefont{A.}~\bibnamefont{Gorsky}}, \bibinfo {author}
  {\bibfnamefont{P.~N.}\ \bibnamefont{Kopnin}},\ and\ \bibinfo {author}
  {\bibfnamefont{A.}~\bibnamefont{Krikun}}}%
   (\bibinfo {year} {2010}),\
  \Eprint{http://arxiv.org/abs/1012.1478}{arXiv:1012.1478 [hep-ph]}%
  \bibAnnoteFile{NoStop}{Gorsky:2010xf}%
%%CITATION = 1012.1478;%%
\bibitem{Kaczmarek:2004gv}%
  \BibitemOpen
  \bibfield{author}{%
  \bibinfo {author} {\bibfnamefont{O.}~\bibnamefont{Kaczmarek}}, \bibinfo
  {author} {\bibfnamefont{F.}~\bibnamefont{Karsch}}, \bibinfo {author}
  {\bibfnamefont{F.}~\bibnamefont{Zantow}},\ and\ \bibinfo {author}
  {\bibfnamefont{P.}~\bibnamefont{Petreczky}},\ }%
  \bibfield{journal}{%
  \Doi{10.1103/PhysRevD.70.074505}{\bibinfo {journal} {Phys. Rev.}}\ }%
  \textbf{\bibinfo {volume} {D70}},\ \bibinfo {pages} {074505} (\bibinfo {year}
  {2004}),\
  \Eprint{http://arxiv.org/abs/hep-lat/0406036}{arXiv:hep-lat/0406036}%
  \bibAnnoteFile{NoStop}{Kaczmarek:2004gv}%
%%CITATION = HEP-LAT/0406036;%%
\bibitem{Kaczmarek:2005ui}%
  \BibitemOpen
  \bibfield{author}{%
  \bibinfo {author} {\bibfnamefont{O.}~\bibnamefont{Kaczmarek}}\ and\ \bibinfo
  {author} {\bibfnamefont{F.}~\bibnamefont{Zantow}},\ }%
  \bibfield{journal}{%
  \Doi{10.1103/PhysRevD.71.114510}{\bibinfo {journal} {Phys. Rev.}}\ }%
  \textbf{\bibinfo {volume} {D71}},\ \bibinfo {pages} {114510} (\bibinfo {year}
  {2005}),\
  \Eprint{http://arxiv.org/abs/hep-lat/0503017}{arXiv:hep-lat/0503017}%
  \bibAnnoteFile{NoStop}{Kaczmarek:2005ui}%
%%CITATION = HEP-LAT/0503017;%%
\bibitem{Gursoy:2009jd}%
  \BibitemOpen
  \bibfield{author}{%
  \bibinfo {author} {\bibfnamefont{U.}~\bibnamefont{Gursoy}}, \bibinfo {author}
  {\bibfnamefont{E.}~\bibnamefont{Kiritsis}}, \bibinfo {author}
  {\bibfnamefont{L.}~\bibnamefont{Mazzanti}},\ and\ \bibinfo {author}
  {\bibfnamefont{F.}~\bibnamefont{Nitti}},\ }%
  \bibfield{journal}{%
  \Doi{10.1016/j.nuclphysb.2009.05.017}{\bibinfo {journal} {Nucl. Phys.}}\ }%
  \textbf{\bibinfo {volume} {B820}},\ \bibinfo {pages} {148} (\bibinfo {year}
  {2009}),\ \Eprint{http://arxiv.org/abs/0903.2859}{arXiv:0903.2859 [hep-th]}%
  \bibAnnoteFile{NoStop}{Gursoy:2009jd}%
%%CITATION = 0903.2859;%%
\bibitem{Alanen:2009xs}%
  \BibitemOpen
  \bibfield{author}{%
  \bibinfo {author} {\bibfnamefont{J.}~\bibnamefont{Alanen}}, \bibinfo {author}
  {\bibfnamefont{K.}~\bibnamefont{Kajantie}},\ and\ \bibinfo {author}
  {\bibfnamefont{V.}~\bibnamefont{Suur-Uski}},\ }%
  \bibfield{journal}{%
  \Doi{10.1103/PhysRevD.80.126008}{\bibinfo {journal} {Phys. Rev.}}\ }%
  \textbf{\bibinfo {volume} {D80}},\ \bibinfo {pages} {126008} (\bibinfo {year}
  {2009}),\ \Eprint{http://arxiv.org/abs/0911.2114}{arXiv:0911.2114 [hep-ph]}%
  \bibAnnoteFile{NoStop}{Alanen:2009xs}%
%%CITATION = 0911.2114;%%
\bibitem{Gursoy:2010fj}%
  \BibitemOpen
  \bibfield{author}{%
  \bibinfo {author} {\bibfnamefont{U.}~\bibnamefont{Gursoy}}, \bibinfo {author}
  {\bibfnamefont{E.}~\bibnamefont{Kiritsis}}, \bibinfo {author}
  {\bibfnamefont{L.}~\bibnamefont{Mazzanti}}, \bibinfo {author}
  {\bibfnamefont{G.}~\bibnamefont{Michalogiorgakis}},\ and\ \bibinfo {author}
  {\bibfnamefont{F.}~\bibnamefont{Nitti}}}%
   (\bibinfo {year} {2010}),\
  \Eprint{http://arxiv.org/abs/1006.5461}{arXiv:1006.5461 [hep-th]}%
  \bibAnnoteFile{NoStop}{Gursoy:2010fj}%
%%CITATION = 1006.5461;%%
\end{thebibliography}%

\end{document}